\documentstyle[11pt,fullpage,psfig,amssymb,citesort]{article}

\def\Tr{{\rm Tr}}
\newcommand{\ket}[1]{|#1\rangle}

\def\m@th{\mathsurround=0pt }
\def\leftrightarrowfill{$\m@th \mathord\leftarrow \mkern-6mu
 \cleaders\hbox{$\mkern-2mu \mathord- \mkern-2mu$}\hfill
 \mkern-6mu \mathord\rightarrow$}
\def\overleftrightarrow#1{\vbox{\ialign{##\crcr
     \leftrightarrowfill\crcr\noalign{\kern-1pt\nointerlineskip}
     $\hfil\displaystyle{#1}\hfil$\crcr}}}
\begin{document}

\renewcommand{\thefootnote}{\fnsymbol{footnote}}
\begin{titlepage}
\begin{flushright}
UFIFT-HEP-99-1 \\
hep-th/9902145
\end{flushright}

\vskip 1.5cm

\begin{center}
\begin{Large}
{\bf Defining the Force between Separated Sources 
on a Light Front\footnote{Supported in part by
the Department of Energy under grant DE-FG02-97ER-41029}}
\end{Large}

\vskip 2.cm

{\large Joel S. Rozowsky\footnote{E-mail  address: rozowsky@phys.ufl.edu}
 and Charles B. Thorn\footnote{E-mail  address: thorn@phys.ufl.edu}}

\vskip 0.5cm

{\it Institute for Fundamental Theory\\
Department of Physics, University of Florida,
Gainesville, FL 32611}

(\today)

\vskip .5cm
\end{center}

\begin{abstract}
The Newtonian character of gauge theories on a light front requires
that the longitudinal momentum $P^+$, which plays the role of
Newtonian mass, be conserved. This requirement conflicts with the
standard definition of the force between two sources in terms of the
minimal energy of quantum gauge fields in the presence of a quark and
anti-quark pinned to {\it points} separated by a distance $R$. We
propose that, on a light front, the force be defined by minimizing the
energy of gauge fields in the presence of a quark and an anti-quark
pinned to {\it lines} (1-branes) oriented in the longitudinal
direction singled out by the light front and separated by a transverse
distance $R$. Such sources will have a limited 1+1 dimensional
dynamics.  We study this proposal for weak coupling gauge theories by
showing how it leads to the Coulomb force law. For QCD we also show
how asymptotic freedom emerges by evaluating the S-matrix through one
loop for the scattering of a particle in the $N_c$ representation of
color $SU(N_c)$ on a 1-brane by a particle in the $\bar N_c$
representation of color on a parallel 1-brane separated from the first
by a distance $R\ll1/\Lambda_{QCD}$. Potential applications to the
problem of confinement on a light front are discussed.
\end{abstract}

\hskip 1 in

\begin{flushright} 
Copyright 1999 by The American Physical Society
\end{flushright}

\vfill
\end{titlepage}

\section{Introduction}
The description of QCD using light-cone methods has little in common
with the more traditional description using Euclidean path integrals.
In particular, the latter approach admits Wilson's elegant criterion
for confinement, that the functional average of the gauge invariant
Wilson loop $\Tr P\exp{ig\oint_Cdx^\mu A_\mu}$ fall off as $e^{-T_0A}$
as the area, $A$, spanned by $C$ gets large~\cite{wilsonconfine}.
Further, for numerical work the path integral has a natural
discretization by Wilson's lattice gauge theory, which has been very
effectively exploited for finer and finer lattices~\cite{lattice98}.
An appealing feature of lattice gauge theory is the manifest gauge
invariance--there is no need for gauge fixing. Although the lattice
spacing breaks $O(4)$ (Euclidean Lorentz) invariance, a discrete
subgroup of $O(4)$ remains, so that violations of $O(4)$ are
irrelevant in the continuum limit.

In sharp contrast, light-cone quantization is only truly useful in a
completely fixed gauge $A^+\equiv(A^0+A^3)/\sqrt2=0$.  Moreover, the
principle advantage of quantization on a light front is the
possibility of interpreting the resulting quantum system in the
language of non-relativistic quantum
mechanics~\cite{weinbergfront,susskindgal,thornfront}: the manifest
space-time symmetry is a Galilei subgroup of the Poincar\'e group
generated by $P^-\equiv H,P^k,M^{+k},M^{12}$.  A seventh generator
$P^+$ plays the role of total Newtonian mass and is also
conserved. The time in this Newtonian analogue is
$t\equiv(x^0+x^3)/\sqrt2$, and if one passes to an imaginary time
formalism in which $it\to\tau$, one is led to a path integral
formalism which has no easily interpreted relation to that of the
original Euclidean gauge theory. For example, it is not obvious how
Wilson's confinement criterion can be implemented in this approach.

The problem is that the Wilson line refers to curves in coordinate
space, whereas the most effective way to exploit light-front dynamics
is to replace $x^-$ by its conjugate $p^+$. The closest one can get to
coordinate space is the ``mixed'' space $x^+,{\bf x},p^+$. Recall that
an $R\times T$ rectangular Wilson loop oriented in the time direction
has the interpretation $e^{-TE(R)}$ as $T\to\infty$ where $E(R)$ is
the lowest energy of a quark and anti-quark pinned to points separated
by a distance $R$. Here we propose instead that we consider a quark
and anti-quark pinned to parallel lines separated by a distance
$R$. If there is confinement, the lowest energy of this system should
still be $\sim T_0R$ for large $R$, with $T_0$ the string tension.
For light-cone dynamics, we can retain $P^+$ conservation by orienting
the parallel lines in the $x^3$ direction.

As a first semi-classical illustration of how this setup works,
imagine a pair of particles constrained to move on two such parallel
lines and interacting (in 3-space) via a potential $V(|\overrightarrow
r|)$. Then in the center of mass system the Hamiltonian will be
\begin{equation}
H=2\sqrt{m^2+p_3^2}+V(\sqrt{z^2+R^2}).
\end{equation} 
For very large $R$ we can approximate
\begin{equation}
V(\sqrt{z^2+R^2})\approx V(R)+{1\over2}{z^2\over R}V^{\prime}(R)
+\cdots
\end{equation}
which is a stable approximation only for an {\it attractive} 
force $V^\prime>0$.
In that case the ground state energy will be, to a good approximation,
\begin{equation}
2m+V(R)+{1\over2}\sqrt{2V^{\prime}(R)\over mR}\approx2m+V(R),
\end{equation}
yielding the potential $V(R)$, as desired.

In the remainder of this paper we will study this idea in the context
of gauge theories. For convenience we shall take the particles living
on the lines (1-branes) to be Dirac fermions. For brevity, we shall
call these constrained particles {\it branions}.  The gauge fields
will of course live in the $3+1$ dimensional ``bulk'' space-time. In
section~\ref{TheEnergy}, we address the problem of calculating the
energy of a pair of 1-branes separated by a distance $R$.  In the weak
coupling limit some sort of ladder approximation is appropriate. We
study three versions of this approximation: the Bethe-Salpeter
equation in Feynman gauge and light-cone gauge, and a Tamm-Dancoff
approximation~\cite{tammdancoff} to the problem of minimizing the
energy of such a system.  For larger coupling, the ladder
approximation fails, and the only simplification available is to the
planar diagrams of large $N_c$ QCD~\cite{thooftlargen}.  In
section~\ref{Scattering}, we study the ``S-matrix'' for a branion on
one line scattering off an anti-branion on the other through one-loop
in light-cone gauge.  For simplicity we assume the (anti-) branions
couple to large $N_c$ QCD in the bulk and are in the ($\bar N_c$)$N_c$
representation of $SU(N_c)$.  We show in detail how asymptotic freedom
emerges and exhibit all of the light-cone ``$P^+=0$'' divergences
which are shown to disappear if virtual momenta are taken on-shell in
a particular way. We conclude in section~\ref{Conclusion} with some
comments on how our proposal can be put to use in studying quark
confinement using light-front methods.

\section{Calculating the Energy}
\label{TheEnergy}
\setcounter{equation}{0}

Our purpose in studying separated 1-branes is to devise
a light-front friendly method to extract the force
between charged sources  separated by a variable distance
$R$. The proposal is to minimize the system energy $U(R)$
with the constraint that each 1-brane carries a
non-zero charge or color, and to identify $-U^\prime(R)$
with the force, at least in certain regimes.

Because the sources are constrained to lines, rather than
to points, this minimization involves the 1+1 dimensional
quantum dynamics of the sources, and this puts certain
limitations on possible applications.
\begin{itemize}
\item
First of all it gives no useful information about sources that repel
one another. In this case the minimum energy configuration is
that with the sources at opposite ends of their respective 1-branes
yielding a minimum energy independent of $R$. This is no
problem for settling the issue of confinement, since that requires the
energy due to separated sources in an overall color singlet state.
Such sources will attract one another whether or not
confinement is realized in the theory.

\item Even if the sources attract one another, the
physical meaning of the minimal energy as a function
of $R$ can be clouded by the brane dynamics. For example, 
if the branions are massless and the actual interaction energy
falls off only as $1/r$ (i.e. confinement does not occur),
the charge on each brane can spread to a size of order $R$
as $R\to\infty$ so, although the minimal energy might fall off
as $1/R$, the coefficient is reduced compared to that of the
actual potential. In that case the method would
establish the absence of confinement but would not yield
a direct measure of the interaction potential.
Giving the branions a mass would provide a limitation to
the growth of the charge size, so that the asymptotic behavior of the minimal
energy would exactly track that of the actual potential. 
\item
In the context of QCD, we expect confinement to
show up as a linear growth $\sim T_0R$ in the minimal energy at large
$R$. However the brane dynamics would not automatically
ensure (via asymptotic freedom) a valid weak coupling description of 
the opposite limit $R\to0$. That would require that {\it all} relevant
momentum scales be large compared to $\Lambda_{QCD}$: not only $1/R$ 
but also the momentum transfers involved in the binding
dynamics.
To track the force from a perturbative description at small $R$
to confinement at large $R$ we require that the branions
be very heavy: $1/a_b=N_cg^2m\gg\Lambda_{QCD}$. Here $a_b$ is
the Bohr radius for the Coulombic bound state, and sets
the size of charge distribution on each brane. If the inequality
holds, the minimal energy will have a renormalization
group improved Coulombic behavior $N_cg^2(R)/4\pi R$ for
$a_b\ll R\ll1/\Lambda_{QCD}$ and, if confinement occurs, a linear growth for
$R\gg1/\Lambda_{QCD}$. The behavior of the minimal energy for
$R\lesssim a_b$ is reflective of the two branion composite system,
and has no direct interpretation as an effective force between
two point sources.
\end{itemize}

Weak coupling methods (valid in QCD for short distance
phenomena) can only effectively describe bound states
close to threshold, that is when the bound state mass
$M(R)\to 2m$ in the zero coupling limit. This is a
nonrelativistic regime: in the center of mass system energy transfers and
the squares of momentum transfers are of order ${\cal O}(2m-M)$.
Because energy transfers are much smaller than momentum transfers,
the propagators of exchanged gluons describe effectively
instantaneous interactions, and the nonrelativistic
Schr\"odinger equation is applicable. This nonrelativistic
dynamics can be identified in quantum field theory in a
number of ways which we shall address below. 
For large distances, where $N_c\alpha_s(R)\sim1$, a
full-fledged nonperturbative treatment is required. 
We advocate an approach based on summing the planar graphs
of 't Hooft's large $N_c$ limit. This limit leads to important 
simplifications, both because it
suppresses pair production and because only 
planar graphs need be summed. Unfortunately,
in the continuum theory no further simplification is possible.
But we think it particularly interesting
to consider a strong  't Hooft coupling limit in the context of
light-cone quantization with discretization of both $P^+$ and $x^+$.
Of course, this limit takes one far from the continuum theory, but it is
likely to be described by string theory~\cite{thornfishnet}.
This possibility has been strengthened recently by 
developments surrounding the Maldacena 
conjecture~\cite{maldacena,gubserkp,wittenholog}.

In the remainder of this section we discuss the energy of two 1-brane
sources in the weak coupling limit of threshold binding. We consider
the configuration of a branion anti-branion pair each confined to
$1+1$-dimensional branes separated by a transverse distance ${\bf R}$.
The gauge fields, which take values in the Lie algebra of $U(N_c)$,
live in the $4$-dimensional bulk space-time.  Our model is therefore
described by the Lagrangian,
\begin{eqnarray}
{\cal L}=-{1\over4}\Tr F_{\mu\nu}F^{\mu\nu}+
\delta({\bf x})\overline{\psi}_1\left[i\gamma^\alpha(\partial_\alpha
-ig A_\alpha)-m\right]\psi_1 \nonumber\\
+\delta({\bf x-R})\overline{\psi}_2\left[i\gamma^\alpha
(\partial_\alpha-ig A_\alpha)-m\right]\psi_2 \,,
\label{Lagrangian}
\end{eqnarray}
where the index $\alpha$ only runs over brane coordinates.
The field strengths are given by $F_{\mu\nu}=\partial_\mu A_\nu
-\partial_\nu A_\mu-ig[A_\mu,A_\nu]$. Note that with this
normalization $\alpha_s=g^2/2\pi$.

It is useful to identify a subset of graphs which is responsible for
binding in this regime as a starting point for understanding what
happens as the coupling is increased.  As $R$ increases from small
values, we can begin to see gradually how more complicated diagrams
become important. Optimistically this inside-out approach may provide
a framework for understanding the dynamics of confinement in the large
$N_c$ limit.  Of course we expect weak coupling binding to be captured
by the sum of ladder diagrams or by the associated Bethe-Salpeter
equation. We first discuss that approach. Ladder diagrams are gauge
dependent, but that dependence should disappear in the
non-relativistic binding regime. We shall confirm this expectation by
examining the ladder approximation in both Feynman and light-cone
gauges.  Later we shall show how Tamm-Dancoff truncation can capture
the same physics.

\subsection{Summing Ladder Feynman Diagrams}
\begin{figure}[ht]
\centerline{\psfig{file=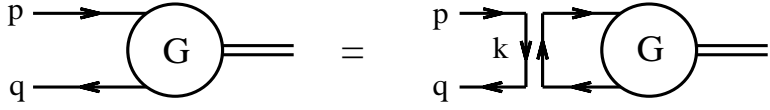,width=4in}}
\caption{Graphical version of the integral equation describing the
ladder sum.}
\label{BetheSalpeter}
\end{figure}
We write the integral equation describing the ladder sum for
the channel where a branion (with 2-momentum $p$) moves on a 
1-brane oriented parallel
to the $z$-axis passing through ${\bf x}=0$ and an anti-branion
(of 2-momentum $q$) moves on a parallel 1-brane 
passing through ${\bf x}={\bf R}$.
Call the Green function, describing the coupling
of the branion and anti-branion to a bound system of 2-momentum
$p+q$, $G(p,q)$. Then
\begin{eqnarray}
G(p,q)=g^2N_c{m+\gamma\cdot q\over m^2+q^2}\int {d{\bf k}\over(2\pi)^2}
e^{i{\bf k}\cdot{\bf R}}{d^2{k}\over(2\pi)^2}
D_{\mu\nu}({\bf k},k)\gamma^\mu G(p-k,q+k)\gamma^\nu
{m-\gamma\cdot p\over m^2+p^2}.
\end{eqnarray}
In this equation Minkowski two-vectors are denoted $p,q,k$, etc.
while vectors in the $xy$-plane are indicated by bold face type.

In Feynman gauge $D_{\mu\nu}=-i\eta_{\mu\nu}/({\bf k}^2+k_3^2-k_0^2)$.
In the non-relativistic binding regime and in the center of mass
system, $k_0\sim (g^2N_c)/R$ while $k_{1,2,3}\sim mg^2N_c$.
Thus, for $R\gg1/m$, $k_0$ can be neglected in the 
denominator of the propagator, and
only $\hat G\equiv \int G dk^0/2\pi$ appears on the r.h.s. Putting
$p^0\to p^0-l^0$, $q^0\to q^0+l^0$, and integrating both sides
with respect to $l^0$ yields by contour integration
\begin{eqnarray}
&&\hat G(p^3,q^3)\approx g^2N_c{m+\gamma^3q^3-\gamma^0\sqrt{m^2+q_3^2}
\over 2\sqrt{m^2+q_3^2}}\int {d{\bf k}\over(2\pi)^2}
e^{i{\bf k}\cdot{\bf R}}{d{k^3}\over(2\pi)}
{1\over{\bf k}^2+k_3^2}\nonumber\\
&&\qquad \gamma_\mu\hat G(p^3-k^3,q^3+k^3)\gamma^\mu
{m-\gamma^3p^3+\gamma^0(p^0+q^0-\sqrt{m^2+q_3^2})\over m^2+p_3^2
-(p^0+q^0-\sqrt{m^2+q_3^2} )^2}\nonumber\\
&&+g^2N_c{m+\gamma^3q^3-\gamma^0(p^0+q^0+\sqrt{m^2+p_3^2})\over m^2+q_3^2
-(p^0+q^0+\sqrt{m^2+p_3^2} )^2}
\int {d{\bf k}\over(2\pi)^2}
e^{i{\bf k}\cdot{\bf R}}{d{k^3}\over(2\pi)}
{1\over{\bf k}^2+k_3^2}\nonumber\\
&&\qquad \gamma_\mu\hat G(p^3-k^3,q^3+k^3)
\gamma^\mu{m-\gamma^3p^3-\gamma^0\sqrt{m^2+p_3^2}
\over 2\sqrt{m^2+p_3^2}}.
\end{eqnarray}
In the center of mass system $q^3=-p^3$ and $q^0+p^0\equiv M$, the
mass of the bound system, is expected to be $2m-{\cal O}(g^2N_c/R)$.
Thus the first term on the r.h.s. is much
larger than the second and, neglecting momenta small compared to $m$,
the equation reduces to  
\begin{eqnarray}
&&\left(2m-M+{q_3^2\over m}\right)\hat G(p^3)\approx g^2N_c
\int{d{k^3}\over2\pi}{d{\bf k}\over(2\pi)^2}
{e^{i{\bf k}\cdot{\bf R}}\over{\bf k}^2+k_3^2}
{1-\gamma^0\over 2} \gamma_\mu \hat G(p^3-k^3)\gamma^\mu
{1+\gamma^0\over 2}.
\end{eqnarray}
This equation implies that $(1+\gamma^0)\hat G=\hat G(1-\gamma^0)=0$,
so that only $\mu=0$ contributes in the contraction of the
gamma matrices. The projection operators can be deleted
and the equation thus reduces to
\begin{eqnarray}
\left(2m-M+{q_3^2\over m}\right)\hat G(p^3) &\approx& g^2N_c
\int{d{k^3}\over2\pi}{d{\bf k}\over(2\pi)^2}
{e^{i{\bf k}\cdot{\bf R}}\over{\bf k}^2+k_3^2}
\hat G(p^3-k^3) \nonumber \\
&\approx&{g^2N_c\over4\pi^2}\int dk^3
K_0(|k^3|R)\hat G(p^3-k^3),
\label{nrbs}
\end{eqnarray}
which is just the momentum space representation of the 
nonrelativistic Schr\"odinger equation
\begin{eqnarray}
\left(-{1\over m}{\partial^2\over\partial z^2}-{N_cg^2\over4\pi
\sqrt{z^2+R^2}}\right)\psi=(M(R)-2m)\psi.
\end{eqnarray}

The reduction is a bit more efficient in light-cone gauge, wherein the
only relevant component of the gluon propagator is
$D^{--}=-2ik^-/k^+({\bf k}^2-2k^+k^-)$. In the nonrelativistic regime
the particles on the branes have energies of order $m+{\cal
O}(N_c\alpha_s/R)$ but momenta of order ${\cal O}(N_c\alpha_sm)$ so
that the momentum transfer $k^-=-k^++{\cal O}(N_c\alpha_s/R)=
{\cal O}(N_c\alpha_s
m)$.  Thus we can make the replacement $D^{--}\to+2i/({\bf
k}^2+2k^{+2})$ in the Bethe-Salpeter equation, which with the
following definition  
$\hat G(p^+,q^+)\equiv\int dl^-\gamma^+G(p-l,q+l)\gamma^+$, 
takes the form,
\begin{eqnarray}
\left[{m^2\over 2p^+}+{m^2\over 2q^+}-(p^-+q^-)\right]\hat G(p,q)
&\approx&2g^2N_c\int {d{\bf k}d{k^+}\over(2\pi)^3}
e^{i{\bf k}\cdot{\bf R}} {\hat G(p-k,q+k)\over {\bf k}^2+2k^{+2}}
\nonumber\\
&\approx&{g^2N_c\over2\pi^2}\int dk^+
K_0(\sqrt2|k^+|R)\hat G(p-k,q+k).
\label{LightConeBetheSalpeter}
\end{eqnarray}
A bound state of mass $M$ would have $p^-+q^-=M^2/2(p^++q^+)$,
so after a little rearrangement, we get
\begin{eqnarray}
{(4m^2-M^2)p^+q^++m^2(p^+-q^+)^2\over 2p^+q^+(p^++q^+)}\hat G(p,q)
\approx{g^2N_c\over2\pi^2}\int dk^+
K_0(\sqrt2|k^+|R)\hat G(p-k,q+k).
\end{eqnarray}
In the nonrelativistic regime, $p^+\approx q^+\approx m/\sqrt2$.
Since $4m^2-M^2\approx(2m-M)4m$ and $(p^+-q^+)^2$ are of the
same order of smallness, it is permissible to make these
substitutions in their coefficients:
\begin{eqnarray}
\left[2m-M+{(p^+-q^+)^2\over2m}\right]\hat G(p,q)
\approx{g^2N_c\over2\sqrt2\pi^2}\int dk^+
K_0(\sqrt2|k^+|R)\hat G(p-k,q+k).
\end{eqnarray}
At fixed $p^++q^+$, $\hat G$ can be regarded as a function of
$q^+-p^+\to p^3\sqrt2$. Thus by changing variables, $k^+\to k^3/\sqrt2$,
we regain Eq.(\ref{nrbs}).

\subsection{Light Front Tamm-Dancoff Approach}
In 2-dimensions the branion Dirac spinors have two components (see
Eq.~(\ref{Lagrangian})), however, in light-cone quantization with
$A_-=0$, they are not independent. Thus, in the operator Hamiltonian
approach one must eliminate one of the components and express the
dynamics in terms of single component spinors. This leads us to the
light-cone Hamiltonian,
\begin{eqnarray}
&&\hskip -1cm {P^-}=
\int dx^-\left[-{im^2\over 2}{\psi}^\dagger_1{1\over\partial_-}\psi_1
-{im^2\over 2}{\psi}^\dagger_2{1\over\partial_-}\psi_2
+{g^2\over 2}\delta({\bf 0}) \Tr\left\{\left({1\over\partial_-} 
{\psi}_1\psi^\dagger_1\right)^2+\left({1\over\partial_-} 
{\psi}_2\psi^\dagger_2\right)^2\right\}\right] \nonumber \\
&&\hskip -0cm
+\int d{\bf x}dx^-\Tr\Biggl[{1\over2}\nabla A^i\cdot\nabla A^i
-g\nabla\cdot{\bf A}\left\{\delta({\bf x}){1\over\partial_-}
\left({\psi}_1\psi^\dagger_1\right) 
+ \delta({\bf x}-{\bf R}) {1\over\partial_-}
\left({\psi}_2\psi^\dagger_2\right)\right\} \nonumber \\
&& \hskip 6cm
+ \mbox{ {\large gluon interaction terms}} 
\hskip .5cm \Biggr],
\label{Hamiltonian}
\end{eqnarray}
where $\psi_1$ and $\psi_2$ are one-component spinors and
$\psi\psi^\dagger$ is understood to be a color matrix
${(\psi\psi^\dagger)_\alpha}^\beta$. Also $\nabla$ is the derivative
operator with respect to the transverse coordinates (transverse to the
1-branes) and the index $i$ runs over transverse components of the
gauge field.  The terms in Eq.~(\ref{Hamiltonian}) proportional to
$\delta({\bf 0})$ are associated with branion dynamics on a single
brane. The divergent coefficients are due to the zero thickness of the
brane, not the UV singularities of the bulk field theory. To deal with
them at higher order, they should be regulated by giving a small
thickness to the branes. In this paper we shall only see the effect of
these terms in the one-loop branion self energy, where it can be
absorbed into mass and wave function renormalization. More generally
when the four branion vertex from these terms is combined with the
corresponding exchange graphs, the most severe quadratic divergence
cancels leaving a net logarithmic divergence.  However, for our
problem of the force between separated branes we do not need to
consider these terms to the order in which we are working.

The fundamental commutation relations of the quantum fields are
implied by their relation to creation and annihilation operators:
\begin{eqnarray}
{{\bf A}_\alpha}^\beta({\bf x},x^-)&=&\int_0^\infty {dk^+\over\sqrt{4\pi k^+}}
\left[{{\bf a}_\alpha}^\beta({\bf x},k^+)e^{-ik^+x^-}+ 
{{{\bf a}^\dagger}_\alpha}^\beta({\bf x},k^+) 
e^{ik^+x^-}\right], \\
{\psi_a}_\alpha(x^-)&=&\int_0^\infty {dp^+\over\sqrt{2\pi}}
\left[{b_a}_\alpha(p^+)e^{-ip^+x^-}+ {d_a^\dagger}_\alpha(p^+) 
e^{ip^+x^-}\right],
\end{eqnarray}
where the index $a$ distinguishes between the branes and
$\alpha$,$\beta$ are color indices.  The creation and annihilation
operators satisfy the usual commutation relations
\begin{eqnarray}
[{{a_k}_\alpha}^\beta({\bf x},p^+),
{{{a_l}^{\!\dagger}}_\gamma}^\delta({\bf y},q^+)]&=&
{\delta_\alpha}^\delta{\delta_\gamma}^\beta
\delta_{kl}\delta({\bf x}-{\bf y}) \delta(p^+-q^+) \nonumber\\
\{{b_r}_\alpha(p^+),{{b_s}^{\!\dagger}}^\beta(q^+)\}&=&
{\delta_\alpha}^\beta\delta_{rs} \delta(p^+-q^+)
\nonumber\\
\{{d_r}^\alpha(p^+),{{d_s}^{\!\dagger}}_\beta(q^+)\}&=& 
{\delta_\beta}^\alpha\delta_{rs}\delta(p^+-q^+).
\end{eqnarray}

We look for an approximate $P^-$ eigenstate of the form
\begin{equation}
\ket{\Psi}={b_1}^\dagger(p^+){d_2}^\dagger(q^+)\ket{0}\phi(p^+,q^+)
+{b_1}^\dagger(p^+){a_j}^\dagger({\bf x},k^+){d_2}^\dagger(q^+)\ket{0}
\chi^j({\bf x},k^+,p^+,q^+),
\end{equation}
where integration over all variables is implicitly understood. When we
apply $P^-$ to $\ket{\Psi}$ we only keep terms proportional to the
same states as in $\ket{\Psi}$ (Tamm-Dancoff
truncation~\cite{tammdancoff}). It is convenient to express
$P^-=P_0^-+P_I^-$, where $P_0^-$ is the non-interacting Hamiltonian
and $P_I^-$ are the terms of ${\cal O}(g)$. Then
\begin{eqnarray}
P^-_0\ket{\Psi}&=&{b_1}^\dagger(p^+){d_2}^\dagger(q^+)\ket{0}
\left({m^2\over2p^+}+{m^2\over2q^+}\right)\phi(p^+,q^+)\nonumber\\
&&+{b_1}^\dagger(p^+){a_j}^\dagger({\bf x},k^+){d_2}^\dagger(q^+)\ket{0}
\left({m^2\over2p^+}+{m^2\over2q^+}-{\nabla^2\over2k^+}\right)
\chi^j({\bf x},k^+,p^+,q^+).
\label{NoInteraction}
\end{eqnarray}

The relevant part of the interaction Hamiltonian can be expressed as
\begin{eqnarray}
P_I^-&=& -g\int d{\bf x}dx^-\;
\Tr\nabla\cdot{\bf A}:\!\left[\delta({\bf x}){1\over\partial_-}
{\psi}_1\psi_1^\dagger + \delta({\bf x}-{\bf R}) {1\over\partial_-}
{\psi}_2\psi_2^\dagger\right]\!: \nonumber \\
&=& -g\int_0^\infty {dp^+dq^+\over\sqrt{4\pi}}
:\!\Tr\Bigg[\left(b_1(q^+)b_1^\dagger(p^+)+d_1^\dagger(p^+) d_1(q^+)\right)
{\nabla\cdot{\bf a}({\bf 0},p^+-q^+)+\nabla\cdot
{\bf a}^\dagger({\bf 0},q^+-p^+) \over i(p^+-q^+)|p^+-q^+|^{1/2}} \nonumber \\
&& \hskip 2.8cm + d_1^\dagger(q^+)b_1^\dagger(p^+) 
{\nabla\cdot{\bf a}({\bf 0},p^++q^+) \over i (p^++q^+)^{3/2}}
- b_1(q^+)d_1(p^+) {\nabla\cdot{\bf a}^\dagger({\bf 0},p^++q^+)
\over i (p^++q^+)^{3/2}} \Bigg]\!: \nonumber \\
&& \hskip 5cm + \hskip 1cm \left[\{1\rightarrow 2\} \hskip .5cm {\rm and} 
\hskip .5cm \{{\bf 0} \rightarrow {\bf R}\}\right],
\end{eqnarray}
where $a_j({\bf x},k^+)=a_j^\dagger({\bf x},k^+)=0$ for $k^+< 0$.
With $P_I^-$ expressed in terms of creation and annihilation operators
one can readily evaluate $P_I^-$ applied to $\ket{\Psi}$. Thus
performing the advertised truncation, we find
\begin{eqnarray}
(P^-_I\ket{\Psi})_{\rm T-D}&=&b_1^\dagger(p^{+\prime})\,
{a_j^\dagger({\bf x},p^+-p^{+\prime})}\,d_2^\dagger(q^+)\ket{0}\,
{ig\nabla^j\delta({\bf x}) \over\sqrt{4\pi}|p^{+\prime}-p^+|^{3/2}}
\phi(p^+,q^+)\nonumber\\
&&-b_1^\dagger(p^{+})\,{a_j^\dagger({\bf x},q^+-p^{+\prime})}\,
d_2^\dagger(p^{+\prime})\ket{0}\,
{ig\nabla^j\delta({\bf x}-{\bf R})\over\sqrt{4\pi}|p^{+\prime}-q^+|^{3/2}}
\phi(p^+,q^+)\nonumber\\
&&+b_1^\dagger(p^{+\prime})\,d_2^\dagger(q^+)\ket{0}\,
{igN_c\delta({\bf x})\over\sqrt{4\pi}|p^{+\prime}-p^+|^{3/2}}
\nabla_j\chi^j({\bf x},p^{+\prime}-p^+,p^+,q^+)\nonumber\\
&&-b_1^\dagger(p^{+})\,d_2^\dagger(p^{+\prime})\ket{0}\,
{igN_c\delta({\bf x}-{\bf R})\over\sqrt{4\pi}|p^{+\prime}-q^+|^{3/2}}
\nabla_j\chi^j({\bf x},p^{+\prime}-q^+,p^+,q^+),
\label{Interaction}
\end{eqnarray}
where the $N_c$ factor arises from a ${\delta_\alpha}^\alpha=N_c$ when
contracting annihilation and creation operators.

We would like to solve the following bound-state energy equation:
\begin{equation}
P^-\ket{\Psi} = \left(P^-_0+P^-_I\right)\ket{\Psi} = E\ket{\Psi}.
\label{EnergyEquation}
\end{equation}
Substituting Eqns~(\ref{NoInteraction}) and (\ref{Interaction}) into
Eq.~(\ref{EnergyEquation}) yields two equations corresponding to each
of the independent states in $\ket{\Psi}$. These are
\begin{eqnarray}
\left({m^2\over2p^+}+{m^2\over2q^+}-E\right)\phi(p^+,q^+) &=& 
-{igN_c\delta({\bf x})\over \sqrt{4\pi}|p^{+\prime}-p^+|^{3/2}}
\nabla_j\chi^j({\bf x},p^+-p^{+\prime},p^{+\prime},q^+)\nonumber\\
&&+{igN_c\delta({\bf x}-{\bf R})\over\sqrt{4\pi}|p^{+\prime}-q^+|^{3/2}}
\nabla_j\chi^j({\bf x},q^+-p^{+\prime},p^+,p^{+\prime}),
\label{EquationforPhiandChi}
\end{eqnarray}
and
\begin{eqnarray}
\left({m^2\over2p^+}+{m^2\over2q^+}-{\nabla^2\over2k^+}-E\right)
\chi^j({\bf x},k^+,p^+,q^+) &=&
-{ig\nabla^j\delta({\bf x})\over\sqrt{4\pi}|k^+|^{3/2}}
\phi(p^++k^{+},q^+)\nonumber\\
&& +{ig\nabla^j\delta({\bf x}-{\bf R})\over\sqrt{4\pi}|k^+|^{3/2}}
\phi(p^{+},q^++k^+).
\label{EquationforChiandPhi}
\end{eqnarray}
We can use Eq.~(\ref{EquationforChiandPhi}) to solve for $\nabla_j\chi^j$
in terms of $\phi$, 
\begin{eqnarray}
\nabla_j\chi^j({\bf x},k^+,p^+,q^+)&=&
\left({\bf x}\left|{-\nabla^2\over-{\nabla^2}+k^+({m^2/p^+}+
{m^2/q^+}-2E)}\right|{\bf 0}\right){ig\over\sqrt{\pi|k^+|}}
\phi(p^++k^{+},q^+)\nonumber\\
&& \hskip -1.2cm
-\left({\bf x}\left|{-\nabla^2\over-{\nabla^2}+k^+({m^2/p^+}+
{m^2/q^+}-2E)}\right|{\bf R}\right){ig\over\sqrt{\pi|k^+|}}
\phi(p^{+},q^++k^+).
\label{ChiintermsofPhi}
\end{eqnarray}
Using Eq.~(\ref{ChiintermsofPhi}) we can eliminate $\nabla_j\chi^j$ from
Eq.~(\ref{EquationforPhiandChi}), yielding
\begin{eqnarray}
&&\left({m^2\over2p^+}+{m^2\over2q^+}-E\right)\phi(p^+,q^+) = \nonumber \\
&& \hskip 2cm \int_0^{p^+}dp^{+\prime}{g^2N_c\over2\pi|p^{+\prime}-p^+|^2}
\Bigg[
\left({\bf 0}\left|{-\nabla^2\over-{\nabla^2}+M^2(p^+,q^+)}
\right|{\bf 0}\right)\phi(p^+,q^+) \nonumber \\
&& \hskip 6cm -\left({\bf 0}\left|{-\nabla^2\over-{\nabla^2}
+M^2(p^+,q^+)}\right|{\bf R}\right)\phi(p^{+\prime},p^++q^+-p^{+\prime})
\Bigg] \nonumber \\
&& \hskip 2cm -\int_0^{q^+}dp^{+\prime}{g^2N_c\over2\pi|p^{+\prime}-q^+|^2}
\Bigg[
\left({\bf R}\left|{-\nabla^2\over-{\nabla^2}+M^2(q^+,p^+)}
\right|{\bf 0}\right)\phi(p^++q^+-p^{+\prime},p^{+\prime}) \nonumber \\
&& \hskip 6cm -\left({\bf R}\left|{-\nabla^2\over-{\nabla^2}
+M^2(q^+,p^+)}\right|{\bf R}\right)\phi(p^+,q^+)\Bigg],
\label{EquationforPhi}
\end{eqnarray}
where 
\begin{equation}
M^2(p^+,q^+) = (p^+-p^{+\prime})\left({m^2\over p^{+\prime}}
+{m^2\over q^+}-2E\right) \,.
\end{equation}
Note in Eq.~(\ref{EquationforPhi}) we have restored the explicit
integration over $p^{+\prime}$.

The inverse kernel of the form $({\bf x}|\cdots|{\bf y})$ is readily
evaluated as
\begin{eqnarray}
\left({\bf R}\left|{-\nabla^2\over-{\nabla^2}+M^2}\right|{\bf 0}\right)
&=&\delta({\bf R})-\left({\bf R}\left|{M^2\over-{\nabla^2}+M^2}
\right|{\bf 0}\right)\nonumber\\
&=&\delta({\bf R})-{M^2\over2\pi}K_0(MR) \,,
\label{InverseKernel}
\end{eqnarray}
where for $R\neq0\Rightarrow\delta(R)=0$.

Substituting the result of Eq.~(\ref{InverseKernel}) into
Eq.~(\ref{EquationforPhi}) yields,
\begin{eqnarray}
&&\Bigg({m_r^2\over2p^+}+{m_r^2\over2q^+}-E
+{g^2N_c\over4\pi^2}\int_0^{p^+}dp^{+\prime}{M^2(p^+,q^+)\over
|p^{+\prime}-p^+|^2}K_0(M(p^+,q^+)\cdot 0) \nonumber \\
&&\hskip 2.3cm +{g^2N_c\over4\pi^2}\int_0^{q^+}dp^{+\prime}{M^2(q^+,p^+)\over
|p^{+\prime}-q^+|^2}K_0(M(q^+,p^+)\cdot 0)\Bigg)\phi(p^+,q^+)  \nonumber \\
&&={g^2N_c\over4\pi^2}\int_0^{p^+}dp^{+\prime}{M^2(p^+,q^+)\over
|p^{+\prime}-p^+|^2}K_0(M(p^+,q^+)R)\phi(p^{+\prime},p^++q^+-p^{+\prime})
\nonumber \\
&&\hskip 2.3cm +{g^2N_c\over4\pi^2}\int_0^{q^+}dp^{+\prime}{M^2(q^+,p^+)\over
|p^{+\prime}-q^+|^2}K_0(M(q^+,p^+)R)\phi(p^++q^+-p^{+\prime},p^{+\prime}).
\label{SecondEquationforPhi}
\end{eqnarray}
In Eq.~(\ref{SecondEquationforPhi}) we have absorbed the divergent
$\delta(0)$ factors into the renormalized branion mass, $m_r$. This
equation is still ill-defined both because of the $K_0(M\cdot 0)$
terms and because the ${p^+}^\prime$ integrations are divergent at
${p^+}^\prime=p^+,q^+$. These divergences can be associated with the
truncation of terms with two or more gluons in the eigenvalue 
equation~\cite{tammdancoff,brodskyl,lcpositronium}.

One approach to this problem is the {\it ad hoc} replacement
$2E\to{m_r^2/p^+}+{m_r^2/q^+}$ in the expression for $M^2$,
which causes the non-integrable singularities to disappear. The only
justification is that this replacement is formally valid to order
$g^2N_c$, and so might arguably be provided if higher terms are
properly included~\cite{tammdancoff,brodskyl,lcpositronium}.  Making
the replacement gives $M^2(p^+,q^+)\to M^2(p^+)=
m_r^2|p^{+\prime}-p^+|^{2}/ p^+p^{+\prime}$. With this simplification
the factors proportional to $K_0(M\cdot 0)$ have the right momentum
dependence to be absorbed into a further renormalization of the
branion mass, $m_r$. Then the integral equation becomes
\begin{eqnarray}
\left({m_r^2\over2p^+}+{m_r^2\over2q^+}-E\right)
\phi(p^+,q^+) &=& {g^2N_c\over4\pi^2}{m_r^2\over p^+}
\int_0^{p^{+}}{dp^{+\prime}\over p^{+\prime}} K_0(M(p^+)R) 
\phi(p^{+\prime},p^++q^+-p^{+\prime})\nonumber\\
&&\hskip -.8cm +{g^2N_c\over4\pi^2}{m_r^2\over q^+}
\int_0^{q^{+}}{dp^{+\prime}\over
p^{+\prime}} K_0(M(q^+)R) \phi(p^++q^+-p^{+\prime},p^{+\prime}).
\end{eqnarray}
If we perform a shift in the integration variable in each integral
this becomes
\begin{eqnarray}
&&\left({m_r^2\over2p^+}+{m_r^2\over2q^+}-E\right)
\phi(p^+,q^+) = \nonumber \\
&& \hskip 2cm {g^2N_c\over4\pi^2}{m_r^2\over p^+}
\int_0^{p^{+}}{dk^{+}\over p^+-k^+} K_0\left({m_r|k^+|R\over 
\sqrt{p^+(p^+-k^+)}}\right) \phi(p^+-k^+,q^++k^+) \nonumber\\
&& \hskip 2cm +{g^2N_c\over4\pi^2}{m_r^2\over q^+}
\int_{-q^+}^0{dk^+\over
q^++k^+} K_0\left({m_r|k^+|R\over \sqrt{q^+(q^++k^+)}}\right)  
\phi(p^+-k^+,q^++k^+).
\label{EquationforPhi2}
\end{eqnarray}

We can show that this reduces to the nonrelativistic Schr\"odinger
equation, Eq.~(\ref{nrbs}).  First we identify $E=p^-+q^-$. In the
nonrelativistic limit $p^+\approx q^+ \approx m/\sqrt{2}$. We also
recognize that the integrals on the r.h.s. of
Eq.~(\ref{EquationforPhi2}) are dominated by the Bessel function,
$K_0$, near $k^+\approx 0$. Thus in the nonrelativistic limit
Eq.~(\ref{EquationforPhi2}) reduces to
\begin{equation}
\left[{m_r^2\over2p^+}+{m_r^2\over2q^+}-(p^-+q^-)\right]
\phi(p^+,q^+) \approx {g^2N_c\over2\pi^2}
\int dk^{+} K_0 (\sqrt{2}k^+R) \phi(p^+-k^+,q^++k^+),
\end{equation} 
which is identical to Eq.~(\ref{LightConeBetheSalpeter}). 

\section{Scattering to One Loop}
\label{Scattering}
\setcounter{equation}{0}

In this section we show asymptotic freedom for the configuration of a
branion anti-branion pair confined to 1-branes separated by a distance
{\bf R}. We shall show this by calculating the four-point scattering
amplitude. For simplicity we shall only calculate the planar diagrams
of 't Hooft's large $N_c$ limit~\cite{thooftlargen}.

We shall work in light-cone gauge ($A_-=0$), and use the following
conventions:
\begin{equation}
\gamma^\pm=(\gamma^0\pm\gamma^3)/\sqrt{2} \hskip 1cm , \hskip 1cm
\{\gamma^\mu,\gamma^\nu\}=-2\eta^{\mu\nu}.
\end{equation} 
Also the space-time metric is given by $\eta^{\mu\nu}= {\rm
diag}[-1,1,1,1]$. We shall also adopt the following notation; lower
case momenta are 2-vectors on the brane while upper case momenta are
bulk 4-vectors. So for a 4-vector $Q$ we distinguish its components
longitudinal to the brane via $Q_\|$ (a 2-vector with light-cone
components $Q^+$, $Q^-$) from its transverse components ${\bf Q}$ (in
{\bf bold-face} - a Euclidean 2-vector).

When we compute the individual Feynman diagrams for the one-loop
four-point ``S-matrix'' it will be convenient to express them in
terms of a Fourier integral
\begin{equation}
\tilde\Gamma(p,q,Q_\|,{\bf R}) = \int {d{\bf Q}\over (2\pi)^2}
\,e^{i{\bf Q}\cdot{\bf R}} \,\Gamma(p,q,Q_\|,{\bf Q}),
\label{GammaEquation}
\end{equation} 
where we shall be calculating $\Gamma(p,q,Q_\|,{\bf Q})$, which has
the explicit ${\bf Q}$ dependence.

\begin{figure}[ht]
\begin{center}
\begin{tabular}{|c|c|}
\hline
\multicolumn{2}{|c|}{\bf Light-Cone Feynman Rules} \\
\hline
&\\[-.25cm]
$
\begin{array}[c]{c}
\psfig{file=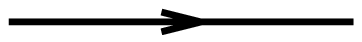,width=0.45in} 
\end{array} 
$
& $-{i\over \gamma^\alpha p_\alpha+m}$ \\[.25cm]
\hline
&\\[-.25cm]
$
\begin{array}[c]{c}
\psfig{file=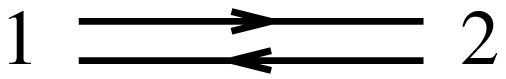,width=0.55in} 
\end{array} 
$
& $ -{i\over K^2}\left(\eta^{\mu_1\mu_2}-{K^{\mu_1}\eta^{\mu_2+}
+K^{\mu_2}\eta^{\mu_1+} \over K^+} \right)$ \\[.25cm]
\hline
&\\[-.25cm]
$
\begin{array}[c]{c}
\psfig{file=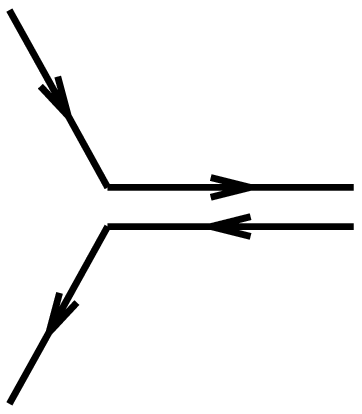,width=0.45in} 
\end{array} 
$
& $ ig\gamma^{\alpha} $ \\
\hline
&\\[-.25cm]
$
\begin{array}[c]{c}
\psfig{file=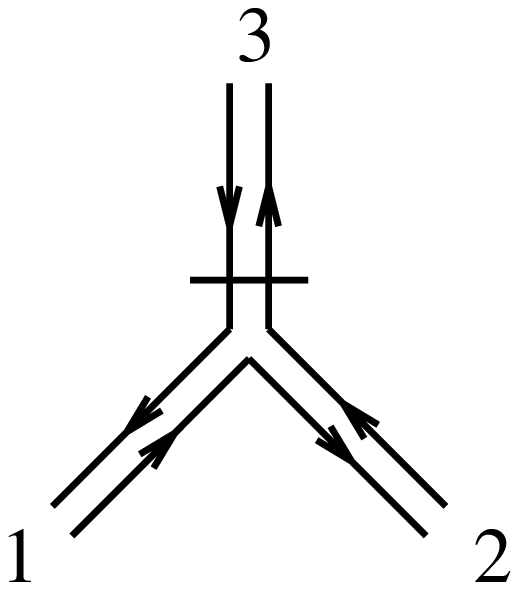,width=0.5in} 
\end{array}
$
& $ -ig\,\eta^{\mu_1 \mu_2}(Q_1-Q_2)^{\mu_3} $ \\
\hline
&\\[-.25cm]
$
\begin{array}[c]{c}
\psfig{file=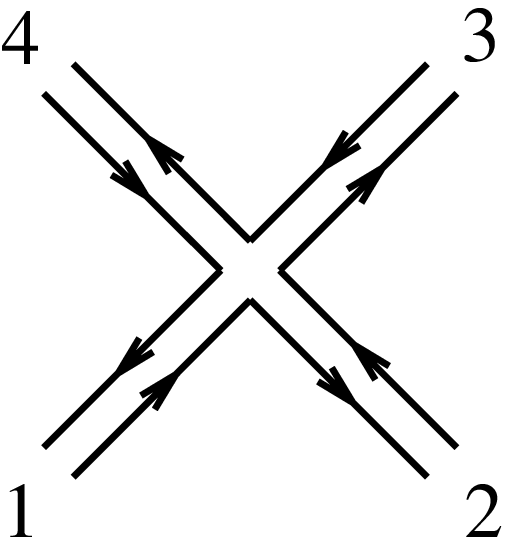,width=0.5in} 
\end{array} 
$
& $ ig^2\left[2\eta^{\mu_1 \mu_3}\eta^{\mu_2 \mu_4}
-\eta^{\mu_1 \mu_2}\eta^{\mu_3 \mu_4}
-\eta^{\mu_1 \mu_4}\eta^{\mu_2 \mu_3}\right] $ \\
\hline
\end{tabular}
\end{center}
\caption{Light-cone Feynman rules using ``double line'' notation. All
momenta in vertices are taken to be incoming and the line in the
three-gluon vertex distinguishes the three cyclic orderings. Index
$\alpha$ only includes brane coordinates, while indices $\mu_i$ run
over all coordinates. }
\label{FeynmanRules}
\end{figure}

With these conventions the light-cone Feynman rules (using the
``double line'' notation) are presented in Fig.\ref{FeynmanRules}. In
the following we shall regulate the integration by an ultraviolet
cutoff on the transverse momenta, ${\bf K}^2 < \Lambda^2$, and an
infrared cutoff on $K^+$, $K^+ > \epsilon$.

\subsection{Branion Wave Function Renormalization}

The Feynman graph for the self-energy is depicted in
Fig.~\ref{SelfEnergy} below.
\begin{figure}[ht]
\centerline{
\psfig{file=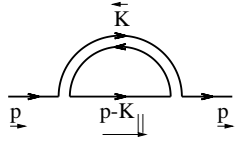,width=1.2in}}
\caption[]{Self Energy graph.}
\label{SelfEnergy}
\end{figure}
Using the light-cone Feynman rules of Fig.~\ref{FeynmanRules} the
self-energy is given by
\begin{equation}
{\partial\Sigma\over\partial p^-} = N_c {\partial\over\partial p^-} 
\int {d^4K\over (2\pi)^4}(ig\gamma^+){-i\over\gamma^\alpha(p+K_\|)_\alpha+m}
(ig\gamma^+)D^{--}(K).
\end{equation}
Thus with some manipulation
\begin{eqnarray}
{\partial\Sigma\over\partial p^-} = {ig^2N_c\gamma^+\over4\pi^2}
\left[\ln{p^+\over\epsilon}\left(\ln{\Lambda^2\over m^2+p^2}-1\right) +
{1\over2}\ln^2{p^+\over\epsilon}+
\int_0^1{dx\over x}\ln{(1-x)(m^2+p^2)\over
(m^2+(1-x)p^2)}\right].
\end{eqnarray}
The branion propagator sandwiched between two vertices proportional to
$\gamma^+$ can be replaced by:
\begin{equation}
-i{\gamma^+(m-\gamma\cdot p)\gamma^+\over m^2+p^2}\to {-ip^+\gamma^+\gamma^-
\gamma^+\over m^2-2p^+p^-}=
{i\gamma^+\over p^--m^2/2p^+},
\end{equation}
from which we can infer that the fermion wave function renormalization is
\begin{equation}
Z_2=1-{g^2N_c\over4\pi^2}
\left[\ln{p^+\over\epsilon}\ln{\Lambda^2\over m^2+p^2} +
{1\over2}\ln^2{p^+\over\epsilon}+ {\rm Finite}\right],
\end{equation}
to one loop. The renormalized four point function will include
a factor of $\sqrt{Z_2}$ for each leg, yielding an overall factor
\begin{eqnarray}
&&1-{g^2N_c\over8\pi^2}
\bigg[\ln{p^+\over\epsilon} \ln{\Lambda^2\over m^2+p^2} +
{1\over2}\ln^2{p^+\over\epsilon}+ \ln{q^+\over\epsilon}
\ln{\Lambda^2\over m^2+q^2} +
{1\over2}\ln^2{q^+\over\epsilon}+\nonumber\\
&&\ln{(p^++Q^+)\over\epsilon}
\ln{\Lambda^2\over m^2+(p+Q_\|)^2} +
{1\over2}\ln^2{(p^++Q^+)\over\epsilon}+\nonumber\\
&&\ln{(q^+-Q^+)\over\epsilon}
\ln{\Lambda^2\over m^2+(q-Q_\|)^2} +
{1\over2}\ln^2{(q^+-Q^+)\over\epsilon}+{\rm Finite}\bigg],
\label{Z2}
\end{eqnarray} 
multiplying the tree amplitude, $A_4^{\rm tree} = 2ig^2\gamma_1^+
\gamma_2^+Q^-/Q^+Q^2$.

\subsection{Gluon Propagator to one Loop}
The one loop corrections to the gluon propagator, shown in
Fig. \ref{GluonPolarization},
 in light-cone gauge
have been evaluated in~\cite{thornfreedom}. After subtraction
of quadratic divergences, the propagator to one loop is given by:
\begin{figure}[ht]
\vskip -1cm
\centerline{
\psfig{file=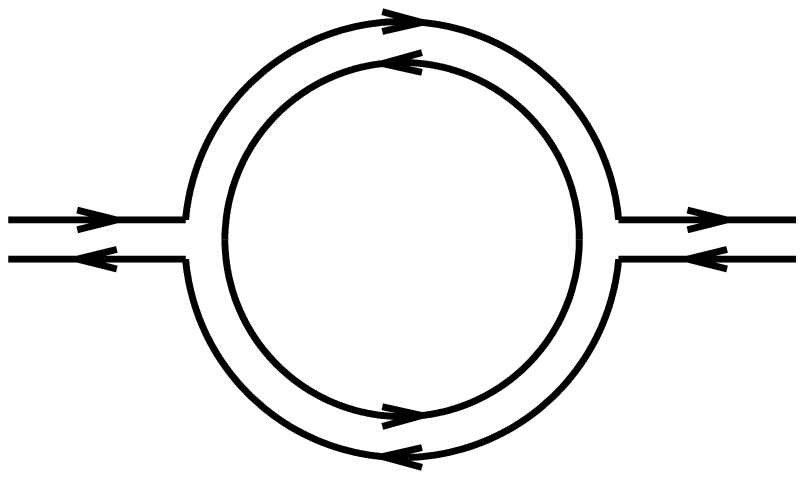,width=1in}
\hskip 2cm
\psfig{file=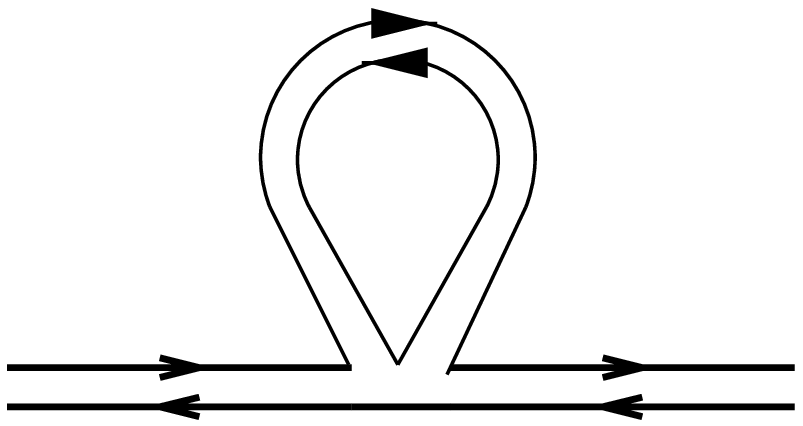,width=1in}}
\caption[]{Feynman diagrams contributing to the one-loop gluon propagator.}
\label{GluonPolarization}
\end{figure}
\begin{eqnarray}
&&D^{--}({\bf Q},Q^+,Q^-)=-{2iQ^-\over Q^+Q^2}
\left[1-{g^2N_c\over16\pi^2}\left\{(8\ln{Q^+\over\epsilon}-{22\over3})
\ln{\Lambda^2\over Q^2}+4\ln^2{Q^+\over\epsilon}
+{4\pi^2\over3}-{134\over9}\right\}\right]\nonumber\\
&&+{i\over Q^{+2}}{g^2N_c\over16\pi^2}\left[8(\ln{Q^+\over\epsilon}-1)
\ln{\Lambda^2\over Q^2}+4\ln^2{Q^+\over\epsilon}
+{4\pi^2\over3}-{206\over9}\right].
\label{GluonPropagator}
\end{eqnarray}

\subsection{Triangle Graph}
\label{TriangleSection}

\begin{figure}[ht]
\centerline{
\psfig{file=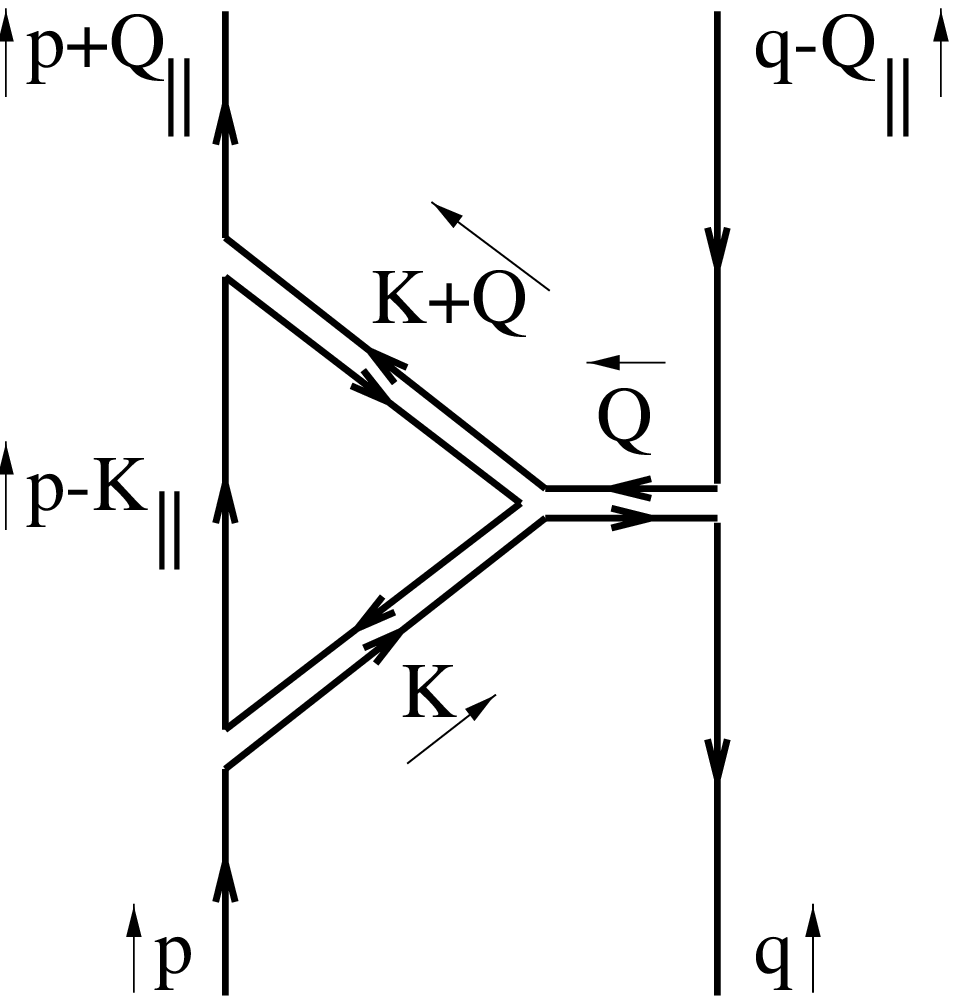,height=4cm}
\hskip 2cm
\psfig{file=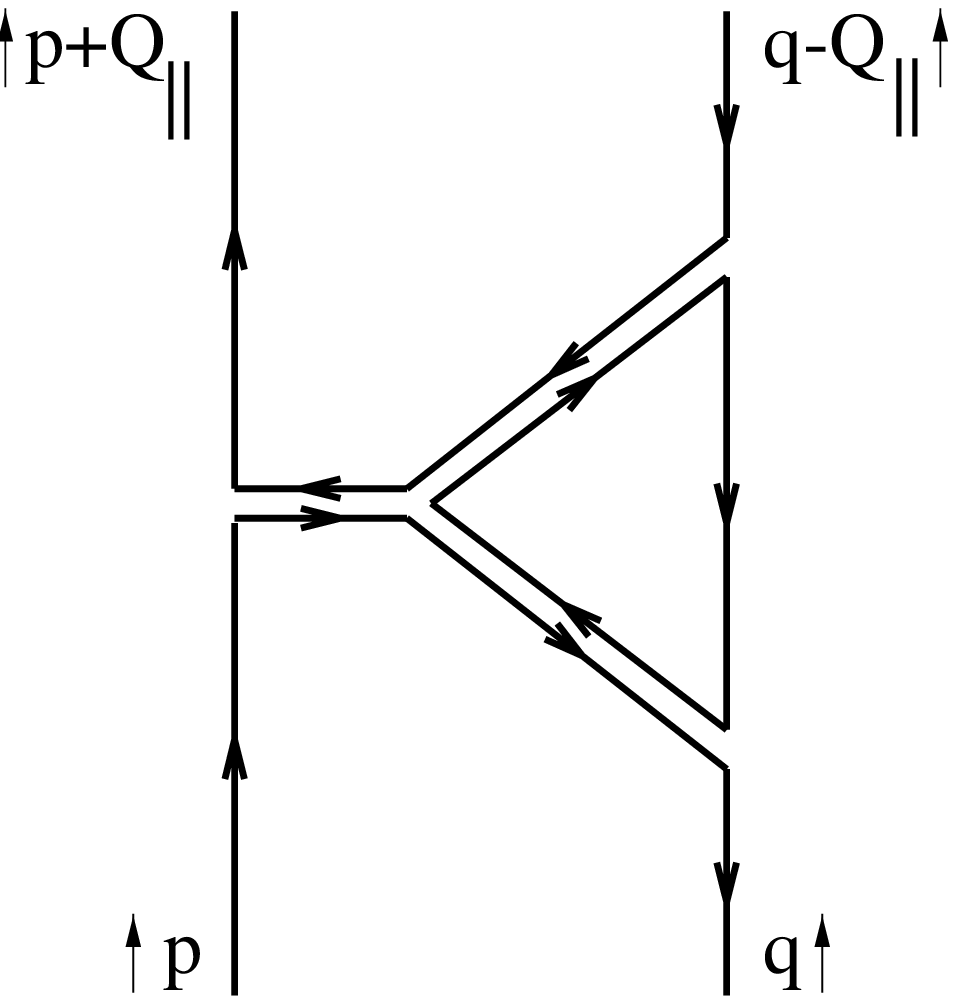,height=4cm}}
\caption[]{Triangle Feynman diagrams contributing to the four-point
amplitude.}
\label{TriangleGraphs}
\end{figure}

We now calculate the triangle graphs contributing to the four-point
amplitude. The Feynman diagrams for these contributions are portrayed
in Fig.~\ref{TriangleGraphs}.
Using the light-cone Feynman rules of Fig.~\ref{FeynmanRules} we
can readily write the Feynman integral corresponding to the diagram on
the l.h.s. of Fig.~\ref{TriangleGraphs}.
\begin{eqnarray}
\Gamma_{\rm Triangle}^{\rm left} &=& N_c \int\!{d^4K\over(2\pi)^4}
(ig\gamma_1^+){-i\over \gamma_1^\alpha(p-K_\|)_\alpha +m}(ig\gamma_1^+)
D^{-\mu_1}(K)D^{-\mu_2}(K+Q)D^{-\mu_3}(Q) \nonumber \\
&& \hskip 3.5cm \times
V_{\mu_1\mu_2\mu_3}(K,-K-Q,Q) (ig\gamma_2^+) \nonumber \\
&=& {4g^4N_c\gamma_1^+\gamma_2^+\over Q^+Q^2} \int\!
{d{\bf K}\over (2\pi)^2}{dK^+ \over 2\pi}{dK^-\over 2\pi}
{(p^+-K^+) \times F \over 
((p\!-\!K_\|)^2+m^2) K^+K^2 (K^+\!+\!Q^+)(K\!+\!Q)^2},
\label{Triangle1}
\end{eqnarray}
where
\begin{equation}
F = K^-[{\bf Q}^2(2Q^+\!+\!K^+)+{\bf K}\cdot{\bf Q}(2K^+\!+\!Q^+)]
-Q^-[{\bf K}^2(2K^+\!+\!Q^+)+{\bf K}\cdot{\bf Q}(2Q^+\!+\!K^+)].
\label{Triangle2}
\end{equation}
The subscripts on the $\gamma$'s distinguish between the different
branes. All branion-branion-gluon vertices only include the $+$
component of $\gamma^\mu$, since the gluon propagator, $D^{\mu\nu}$,
vanishes when $\mu=+$. 

The procedure to evaluate Eq.~(\ref{Triangle1}) is relatively
straightforward. We first perform the $K^-$ integration via contour
integration (inserting the appropriate factors of $i\epsilon$). We are
only interested in obtaining all the log (and ${\rm log}^2$) divergent
terms.  These arise as dependence on the ultraviolet cutoff, $\Lambda$,
and on the infrared cutoff, $\epsilon$. By investigating these
divergences separately and combining, we obtain all the divergent
pieces.

For convenience we introduce the following compact notation 
\begin{equation}
w^* = {w^2 + m^2 \over w^+},
\label{StarNotation}
\end{equation} 
where $w$ is a brane 2-vector such that $w^2=-2w^+w^-$. Thus the
result for the triangle graph on the left-side of
Fig.~\ref{TriangleGraphs} is
\begin{eqnarray}
\Gamma_{\rm Triangle}^{\rm left} &=&
{ig^4N_c\gamma_1^+\gamma_2^+\over4\pi^2Q^2Q^{+2}}\bigg\{
Q^2\left[\left(2\ln{Q^+\over\epsilon}-1\right)
\ln{\Lambda^2\over Q^2}+\ln^2{Q^+\over\epsilon}\right]
\nonumber \\
&&+Q^+Q^-\left[\left(2\ln{Q^+\over\epsilon}+\ln{p^+\over\epsilon}
+\ln{p^++Q^+\over\epsilon}\right)
\ln{\Lambda^2\over Q^2}+2\ln^2{Q^+\over\epsilon}\right]
\nonumber\\
&&\mbox{}
+\ln{Q^+\over\epsilon}\left[
{(Q^2+2Q^+Q^-)(p+Q_\|)^*\over (p+Q_\|)^*-Q^2/Q^+}\ln{Q^2/Q^+\over (p+Q_\|)^*}
\right. \nonumber \\
&& \hskip 3cm \left.
+{(Q^2+2Q^+Q^-)p^*\over p^*+Q^2/Q^+}\ln{Q^2/Q^+\over p^*}\right]
+{\rm Finite}\bigg\}.
\label{TriangleResult}
\end{eqnarray}
A similar result corresponding to the Feynman diagram on the
right-side of Fig.~\ref{TriangleGraphs} may be obtain from
Eq.~(\ref{TriangleResult}) by the substitution, $p\rightarrow q-Q_\|$.

\subsection{Box Graph}
\label{BoxSection}

\begin{figure}[ht]
\centerline{
\psfig{file=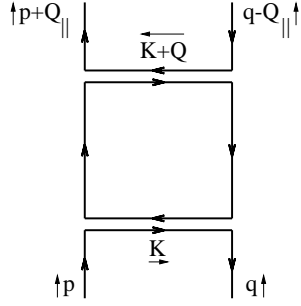,height=4cm}}
\caption[]{Box diagram contributing to the four-point amplitude.}
\label{BoxGraph}
\end{figure}

We proceed to the box graph contributing to the four-point amplitude
which is portrayed in Fig.~\ref{BoxGraph}. 
Thus the Feynman integral corresponding to this is given by
\begin{eqnarray}
&&\Gamma_{\rm Box} = N_c \int {d^4K\over (2\pi)^4}
(ig\gamma_1^+){-i\over \gamma_1^\alpha(p-K_\|)_\alpha +m}(ig\gamma_1^+)
(ig\gamma_2^+){-i\over -\gamma_2^\alpha(Q_\|+K_\|)_\alpha +m}(ig\gamma_2^+)
\nonumber \\
&& \hskip 7cm \times D^{--}(K)D^{--}(K+Q) \\
&&= 16g^4N_c\gamma_1^+\gamma_2^+ \int \!
{d{\bf K}\over (2\pi)^2}{dK^+ \over 2\pi}{dK^-\over 2\pi}
{(p^+\!-\!K^+)(K^+\!+\!q^+)K^-(K^-\!+\!Q^-) \over K^+K^2(K^+\!+\!Q^+)
(K\!+\!Q)^2((p\!-\!K_\|)^2\!+\!m^2)((q\!+\!K_\|)^2\!+\!m^2)}.
\nonumber 
\end{eqnarray}

This integral can be evaluated in a similar manner to the triangle
integral encountered in subsection~\ref{TriangleSection}. We again
first perform the contour integration over $K^-$. We then individually
target the divergences associated with the transverse integration,
$d{\bf K}$ and those associated with the $K^+$ integration. Combining
these contributions yields all the divergent pieces of this
integral. Then again using the notation adopted in
Eq.~(\ref{StarNotation}) we get
\begin{eqnarray}
\Gamma_{\rm Box} &=&
-{ig^4N_c\gamma_1^+\gamma_2^+\over4\pi^2Q^{+2}}\Bigg\{
2\ln{Q^+\over\epsilon}\ln{\Lambda^2\over Q^2}+
\ln^2{Q^+\over\epsilon} +\ln{Q^+\over\epsilon} \Bigg[ \nonumber \\
&& {1 \over (p\!+\!Q_\|)^*\!+\!(q\!-\!Q_\|)^*}
\Bigg( {(q\!-\!Q_\|)^*((q\!-\!Q_\|)^*\!-\!2Q^-)\over 
(q\!-\!Q_\|)^*\!+\!Q^2/Q^+} 
\ln{Q^2/Q^+\over (q\!-\!Q_\|)^*} \nonumber \\
&& \hskip 2cm +{(p\!+\!Q_\|)^*
((p\!+\!Q_\|)^*\!+\!2Q^-)\over (p\!+\!Q_\|)^*\!-\!Q^2/Q^+} 
\ln{Q^2/Q^+\over (p\!+\!Q_\|)^*)}
\Bigg) \nonumber \\
&& + {1 \over p^*\!+\!q^*}
\Bigg( {p^*(p^*\!-\!2Q^-)\over p^*\!+\!Q^2/Q^+} \ln{Q^2/Q^+\over p^*}
+{q^*(q^*\!+\!2Q^-)\over q^*\!-\!Q^2/Q^+} \ln{Q^2/Q^+\over q^*}\Bigg)
\Bigg] +\mbox{ Finite}\Bigg\}.
\label{BoxResult}
\end{eqnarray}

It is noteworthy that after multiplying by $e^{i{\bf Q}\cdot{\bf R}}$
and integrating over ${\bf Q}$ as prescribed in
Eq.~(\ref{GammaEquation}), the ultraviolet divergences $\ln\Lambda^2$ and
the $\ln^2\epsilon$ divergences disappear because they are multiplied
by $\delta({\bf R})=0$ at finite separation. Thus the box is not
needed to obtain asymptotic freedom in this branion scattering
process. The $\ln\epsilon$ divergences, however, remain and are
necessary to eventually obtain $\epsilon$ independent
on-shell scattering.

\subsection{Divergence structure of the One Loop 4 Point Function}
We have assembled the various pieces necessary to construct the
divergent structure of the off-shell one-loop four-point Green
function. The one-loop four-point amplitude is given by: four-point
trees with a factor of $\sqrt{Z_2}$, see Eq.~(\ref{Z2}), for each
external leg; a four-point exchange diagram with the one-loop gluon
propagator, Eq.~(\ref{GluonPropagator}); the triangle graphs of
Fig.~\ref{TriangleGraphs} (the solution for the graph on the right can
be obtained by a simple substitution in the solution for the graph on
the left, Eq.~(\ref{TriangleResult})); the box graph,
Eq.~(\ref{BoxResult}). Since we have taken $N_c\rightarrow\infty$, we
have of course not included non-planar contributions, e.g. from the
crossed box graph.

When we simplify the four-point function all terms proportional to
$\ln \Lambda^2/Q^2$ cancel up to the expected ${11 \over 3}
\ln\Lambda^2/Q^2$ term which is the correct asymptotic behavior for
QCD. Also all terms proportional to $\ln^2 Q^+/\epsilon$ cancel. 
Note that these cancelations occur before integrating
over ${\bf Q}$, but do involve {\it all} of the one loop diagrams. As
mentioned in subsection~\ref{BoxSection} the box diagram
shows no $\ln\Lambda$ or $\ln^2\epsilon$ dependence 
after integrating over ${\bf Q}$ as prescribed
in Eq.~(\ref{GammaEquation}). 

The remaining divergent structure of the off-shell four-point Green
function is given by
\begin{eqnarray}
&& \hskip -1cm {ig^4N_c\gamma^+_1\gamma^+_2 Q^- \over4\pi^2 Q^+Q^2}
\bigg[{11\over3}\ln{\Lambda^2\over Q^2} + {1\over Q^+Q^-}
\ln{Q^+ \over \epsilon}
\bigg\{  { Q^2p^*q^* + Q^2Q^-(p^*\!-\!q^*) +Q^+Q^-p^*(p^*\!+\!q^*) \over
(p^*\!+\!q^*)(p^*\!+\!Q^2/Q^+)} \ln{Q^2/Q^+ \over p^*} \nonumber\\
&& \hskip -1cm +{ Q^2(p\!+\!Q_\|)^*(q\!-\!Q_\|)^*\!+\!Q^2Q^-((q\!-\!Q_\|)^*
\!-\!(p\!+\!Q_\|)^*) 
\!+\! Q^+Q^-(p\!+\!Q_\|)^* ((p\!+\!Q_\|)^*\!+\!(q\!-\!Q_\|)^*) \over
((p\!+\!Q_\|)^*+(q\!-\!Q_\|)^*)((p\!+\!Q_\|)^*-Q^2/Q^+)} 
\ln{Q^2/Q^+ \over (p\!+\!Q_\|)^*} \nonumber\\
&& \hskip -1cm +{ Q^2(p\!+\!Q_\|)^*(q\!-\!Q_\|)^*\!+\!Q^2Q^-((q\!-\!Q_\|)^*
\!-\!(p\!+\!Q_\|)^*) 
\!+\!Q^+Q^-(q\!-\!Q_\|)^* ((p\!+\!Q_\|)^*+(q\!-\!Q_\|)^*) \over
((p\!+\!Q_\|)^*+(q\!-\!Q_\|)^*)((q\!-\!Q_\|)^*+Q^2/Q^+)} 
\ln{Q^2/Q^+ \over (q\!-\!Q_\|)^*} \nonumber\\
&& \hskip -1cm + { Q^2p^*q^* + Q^2Q^-(p^*\!-\!q^*) +Q^+Q^-q^*(p^*\!+\!q^*) 
\over (p^*\!+\!q^*)(q^*\!-\!Q^2/Q^+)} \ln{Q^2/Q^+ \over q^*} 
\bigg\} \; + {\rm Finite} \; \bigg].
\end{eqnarray}

Since this off-shell amplitude is not, as it stands, a physical
quantity, it is not necessary that the $\ln\epsilon$ divergences
cancel.  Even the on-shell limit is not quite physical because of the
usual IR divergences associated with the possibility of soft gluon
bremsstrahlung. In particular the terms we have labeled ``Finite''
are cutoff independent and finite off-shell, but display IR
divergences in the on-shell limit. Focusing on the $\ln\epsilon$
terms, we see that the on-shell limit involves ambiguous terms of the
form $0/0$. Dropping the terms that unambiguously vanish
on-shell, these terms simplify to:
\begin{equation}
{ig^4N_c\gamma^+_1\gamma^+_2 Q^- \over4\pi^2 Q^+Q^2}
\ln{Q^+\over\epsilon} \left[ {p^*-q^*\over p^*+q^*} \ln{q^*\over p^*}
\,+\, {(p\!+\!Q_\|)^*-(q\!-\!Q_\|)^*\over (p\!+\!Q_\|)^*+(q\!-\!Q_\|)^*}
\ln{(q\!-\!Q_\|)^*\over (p\!+\!Q_\|)^*} \right].
\end{equation}
Since $w^* = m^2/w^+-2w^-$ we see that by taking the branions in
either the initial or final states equally off energy shell these
ambiguous terms can be eliminated. Note that this restriction still
allows the total energy to be off-shell, an important flexibility for
bound state problems.  While we are free to make this restriction on
the initial and final states, it would not be allowed if this
off-shell amplitude were part of a larger diagram.  In that case, there
must be other $\ln\epsilon$ terms for them to cancel against.  We have
not yet managed to show that this actually happens.  Even so we have
achieved our goal in this section of demonstrating asymptotic freedom,
which did involve an intricate cancelation of $\ln\epsilon$
divergences in the coefficient of $\ln\Lambda^2$.

\section{Discussion and Conclusion}
\label{Conclusion}
\setcounter{equation}{0}
In this paper we have proposed a light-front friendly way to extract
the interaction energy between two sources in a gauge theory. We have
analyzed the method in several ways at weak coupling.  But the larger
motivation for this proposal is to develop a framework within a
light-cone quantization approach for establishing quark confinement in
non-abelian gauge theories.

Consider, for example, the 't Hooft limit of $N_c\to\infty$, which is
well known to reduce to summing planar graphs. Another version of this
limit interprets the sum of planar diagrams on a light front as the
quantum dynamics of a gluonic chain~\cite{thornfock}.  Without the
notion of pinned sources, these chains would be dynamical bound
states, and the concept of a confining force would have to be
indirectly inferred from the excitation spectrum of these chains. In
ordinary equal time quantization, the concept of a Wilson loop allows
a direct definition of the confining force, however the possibility of
vacuum fluctuations clouds the chain interpretation of the large $N_c$
limit. Our goal was to modify the Wilson criterion to make it suitable
for light front physics.

In light-cone quantization the sum of planar diagrams coupled to two
separated 1-branes can be interpreted as the dynamics of a
nonrelativistic chain of gluons vibrating in the transverse space and
stretched between the points in transverse space marking the locations
of the 1-branes. Since the $p^+$ of each gluon and branion is
dynamical, the Newtonian mass of each of these objects varies; but in
such a way that the total mass is conserved~\cite{thornfront}. This
picture is made more concrete when the $p^+$ is discretized, $p^+\to
Mm$, where $M$ is a large integer, and $m$ is the discrete unit of
$p^+$. We hope that pinning the ends of the chain to points will offer
conceptual and technical simplifications to the problem of assessing
whether the dynamics of the chain leads to a confining force.

Although the dynamics of continuum gauge theories generically involves
a scale dependent coupling which is presumably never actually much
larger than unity, the introduction of an ultraviolet cutoff allows a
formal strong coupling limit. For example, the discretization of $x^+$
(in addition to $p^+$) provides such a cutoff~\cite{thornfishnet}. In
that context the strong 't Hooft coupling limit favors large planar
fishnet diagrams, which should behave as seamless world sheets. In
this limit then the sum of planar diagrams coupled to our 1-branes
would correspond to a fundamental light-cone string stretched between
them, and therefore a confining force would emerge. However, the
precise nature of such a ``fishnet'' approximation to QCD is yet to be
determined.

\underline{Acknowledgments:} We would like to thank K. Bering,
S. Brodsky, S. Glazek, A. Mueller, H. Neuberger, and P. Ramond for
useful discussions.

\end{document}